\documentclass[12pt]{article}
\usepackage{amsfonts,amssymb}
\makeatletter 
\def\EquationsBySection{\def\theequation
{\thesection.\arabic{equation}}%
\@addtoreset{equation}{section}}

\newcommand\old[1]{}
\newtheorem{theorem}{Theorem}[section]
\newtheorem{definition}[theorem]{Definition}
\newtheorem{lemma}[theorem]{Lemma}

\newtheorem{example}[theorem]{Ex}

\EquationsBySection \makeatother

\begin{document}
\pagestyle{plain}
\title
{\bf Degree-distribution Stability of Evolving
Networks\thanks{Supported by National Natural Science Foundation of
China No.10671212, and Research Fund for the Doctoral Program of
Higher Education of China No.20050533036. }}

\author{Zhenting Hou$^{1}$\thanks{Email:
zthou@csu.edu.cn}, Xiangxing Kong$^{1}$, Dinghua Shi$^{2}$, Guanrong
Chen$^{3}$,\\ Qinggui Zhao$^{1}$\\
$^{1}$ School of Mathematics, Central South University,\\
 Changsha, Hunan, 410075, China\\
$^{2}$ Department of Mathematics, Shanghai University,\\
 Shanghai 200444, China\\
$^{3}$ Department of Electronic Engineering, City University of Hong Kong,\\
 Hong Kong, China
}

\date{}
\maketitle

{\bf Abstract:} In this paper, we study a class of stochastic
processes, called evolving network Markov chains, in evolving
networks. Our approach is to transform the degree distribution
problem of an evolving network to a corresponding problem of
evolving network Markov chains. We investigate the evolving network
Markov chains, thereby obtaining some exact formulas as well as a
precise criterion for determining whether the steady degree
distribution of the evolving network is a power-law or not. With
this new method, we finally obtain a rigorous, exact and unified
solution of the steady degree distribution of the evolving network.


%\newpage

\section{Introduction}

\indent Barab\'asi and Albert (BA) found$^{[1]}$ that for many
real-world networks, the fraction of nodes with degree $k$ is
proportional over a large range to a power-law tail, i.e., $P(k)\sim
k^{-\gamma}$, where $\gamma$ is a constant independent of the size
of the network. For the purpose of establishing a mechanism to
produce scale-free properties, they proposed the now-well-known BA
model based on growth and preferential attachment. However, many
real networks are not purely growing (as BA model), instead they are
evolving networks with adding and also removing links and nodes
throughout the developing process. A typical example is the
protein-protein network, which has gene duplication, divergence,
deletion, and heterodimerization.

\indent There are many empirical and simulation studies on evolving
networks$^{[2,3]}$, but analytical models are rare. To name one, Shi
et al.$^{[4]}$ proposed a birth-and-death processing method to
compute the degree distribution of an evolving network. In this
paper, we study a class of stochastic processes, called evolving
network Markov chains, on evolving networks. Our approach is to
transform the degree distribution problem of an evolving network to
a corresponding problem of evolving network Markov chains. We
investigate the evolving network Markov chains, thereby obtaining
some exact formulas as well as a precise criterion for determining
whether the steady degree distribution of the evolving network is a
power-law or not. With this new method, we finally obtain a
rigorous, exact and unified solution of the steady degree
distribution of the evolving network. In another recent
work$^{[5]}$, we have carried out the same, but for growing networks
instead.

\section{ Main Results}

\indent For any $i=1,2,\cdots\,$, Let $k_i(t)$ $(t=i,i+1,\cdots)$ be
a Markov chain taking values in $\{0,1,2,\cdots\,\}$, with initial
distribution $P\{k_i(i)=k\}=d_{k,i}$ and the transition probability
\begin{eqnarray}
 P\{k_i(t+1)=l\,|\,k_i(t)=k\}=\left\{\begin{array}{ll}
 f_t^+(k), &\textrm{$l=k+1$}\\
 f_t^-(k), &\textrm{$l=k-1$}\\
 1-f_t^+(k)-f_t^-(k), &\textrm{$l=k$}\\
 0, &\textrm{else}
\end{array}\right.
\end{eqnarray}
where $0<f_t^+(k)$, $0<f_t^-(k)$, $f_t^+(k)+f_t^-(k)\leq 1$, and
$f_t^-(0):=0$.

\indent Denote $P(k,i,t):=P\{k_i(t)=k\}$ $(t=i,i+1,\cdots$) and
$P(k,t):=\frac{1}{t}\sum\limits_{i=1}^t\,P(k,i,t)$.

\begin{definition}\label{de:1}
The Markov chain $\{k_i(t)\}_{t=i,i+1,\cdots}$ $(i=1,2,\cdots)$ is
called a series of evolving network Markov chains, or simply,
evolving network Markov chains, if the limit
$P(k):=\lim\limits_{t\rightarrow\infty}P(k,t)$ exists and
\begin{eqnarray}
 P(k)\geq0,\quad \sum\limits_{k=0}^\infty P(k)=1
\end{eqnarray}
In this case, it is said that the degree distribution of evolving
network Markov chains exists, and $P(k)$ is called the steady degree
distribution of $\{k_i(t)\}$. Further, if $P(k)$ is power-law, i.e.,
\begin{eqnarray}
 P(k)\sim k^{-\gamma}\quad(\gamma>1)
\end{eqnarray}
then $\{k_i(t)\}$ are called scale-free evolving network Markov
chains.
\end{definition}
\bigbreak

{\bf Assumptions}

\indent (I) The limits $\lim\limits_{t\rightarrow\infty}tf_t^+(k)$
and $\lim\limits_{t\rightarrow\infty}tf_t^-(k)$ exist, denoted by
$F^+(k)$ and $F^-(k)$, respectively.

\indent (II) The limits $\lim\limits_{t\rightarrow\infty}P(k,t),
k=0,1,2,\cdots$ exist. \bigbreak

\indent {\bf Note.}\ Assumption (I) is always satisfied for all the
existing network models. Assumption (II) is also always satisfied
for growing networks. \bigbreak

\begin{theorem}\label{th:1}
If $d_k:=\lim\limits_{t\rightarrow\infty}d_{k,t}$ exists and
satisfies $\sum\limits_{k=0}^\infty d_k=1$, then the following
relations are satisfied for $P(k)$, $k=0,1,2,\cdots\,$:
\begin{eqnarray}\label{th:1.1}
 P(k)=\left\{\begin{array}{ll}
 \frac{F^-(1)}{1+F^+(0)}P(k+1)+\frac{d_0}{1+F^+(0)}, &\textrm{$k=0$}\\
 \frac{F^+(k-1)}{1+F^+(k)+F^-(k)}P(k-1)
 +\frac{F^-(k+1)}{1+F^+(k)+F^-(k)}P(k+1)+\frac{d_{k}}{1+F^+(k)+F^-(k)},
 &\textrm{$k>0$}\\
\end{array}\right.
\end{eqnarray}
\bigbreak

\indent Further, if there are constants $A,B,\overline{A},
\overline{B}$, satisfying $F^+(k)=Ak+B,F^-(k)=
\overline{A}k+\overline{B}$, then
\begin{eqnarray}\label{th:1.2}
 P(k)=\left\{\begin{array}{ll}
 \frac{\overline{A}+\overline{B}}{1+B}P(k+1)+\frac{d_0}{1+B},
 &\textrm{$k=0$}\\
 \frac{A(k-1)+B}{1+Ak+B+\overline{A}k+\overline{B}}P(k-1)
 +\frac{\overline{A}(k+1)+\overline{B}}{1+Ak+B+\overline{A}k
 +\overline{B}}P(k+1)+\frac{d_k}{1+Ak+B+\overline{A}k+\overline{B}},
 &\textrm{$k>0$}\\
\end{array}\right.
\end{eqnarray}
\end{theorem}
\bigbreak

\indent{\bf Note.}\ Due to the preferential attachment, one has
$A\geq 0$ and $\overline{A}\geq 0$, and moreover $A$ and
$\overline{A}$ are not both 0. In addition, $B\geq 0$ since there is
a possibility for a node to receive new links. And the probability
that a node of degree 0 loses a link is 0. Thus, $f_t^-(k)=0$, and
moreover $F^-(0)=0$. Also, the probability that a node of degree 1
loses a link is non-negative, therefore $\overline{A}
+\overline{B}\geq0$. \bigbreak

\begin{theorem}\label{th:2}
Suppose that $(i)$ when $F^+(k)=Ak+B$,
$F^-(k)=\overline{A}k+\overline{B}$, and there are $0\leq m \leq M
<\infty $ such that $d_k=0$ when $k< m$ or $k>M$, the degree
distribution of evolving network Markov chains satisfies
\begin{eqnarray}\label{th:2.1}
P(k)=\left\{\begin{array}{ll}
 \lim\limits_{\varepsilon\rightarrow0}\frac{-\int_\varepsilon^{h}
 b_2(s)e^{-\int_\varepsilon^sa(\theta)d\theta}ds}{1+\overline{B}
 \int_\varepsilon^hb_1(s)e^{-\int_\varepsilon^sa(\theta)d\theta}ds}>0,
 &\textrm{$k=0$}\\[5pt]
 \frac{det D_j}{g\sum\limits_{i=1}^{M}e_i}\,, &\textrm{$1\leq k\leq M-1$}\\
\end{array}\right.
\end{eqnarray}
$(ii)$ when $k\geq M$, the degree distribution of evolving network
Markov chains satisfies
\begin{eqnarray}\label{th:2.2}
P(k)=\left\{\begin{array}{ll}
 C\int_0^1z^{k-1+\frac{1+\overline{B}}{\overline{A}}}(1-z)^{-\frac{1}
 {\overline{A}}}e^{\frac{B}{\overline{A}}\frac{1}{z}}dz,
 &\textrm{$A=0,\overline{A}\neq0$}\\
 C\int_0^1z^{k-1+\frac{B}{A}}(1-z)^{\frac{1}{A}}e^{-\frac{\overline{B}}
 {A}z}dz, &\textrm{$\overline{A}=0,A\neq0$}\\
 C\int_0^1z^{k-1+\frac{B}{A}}(1-z)^{\frac{\overline{B}-B}{A}}
 e^{\frac{1}{A}\frac{1}{z-1}}dz, &\textrm{$\overline{A}=A\neq0$}\\
 C\int_0^1z^{k-1+\frac{B}{A}}(1-z)^{\frac{1}{A-\overline{A}}}
 \left|z-\frac{A}{\overline{A}}\right|^{\frac{\overline{B}}{\overline{A}}
 -\frac{1}{A-\overline{A}}-\frac{B}{A}}dz, &\textrm{else}\\
\end{array}\right.
\end{eqnarray}
where $\varepsilon>0$ is small, and
\begin{eqnarray}\label{th:2.3}
 h=\left\{\begin{array}{ll}
 1, &\textrm{$A\leq\overline{A}$}\\
 \frac{\overline{A}}{A},& \textrm{$A>\overline{A}$ and $\frac{B}{A}
 +\frac{1}{A-\overline{A}}-\frac{\overline{B}}{\overline{A}}>0$}\\
 \varepsilon^2, &\textrm{$A>\overline{A}$ and $\frac{B}{A}
 +\frac{1}{A-\overline{A}}-\frac{\overline{B}}{\overline{A}}\leq0$}\\
\end{array}\right.
\end{eqnarray}
\begin{eqnarray}\label{th:2.4}
 a(z)&=&-\frac{1}{A}\frac{Bz^2-(1+B+\overline{B})z+\overline{B}}
 {z(1-z)(\frac{\overline{A}}{A}-z)}\\
 b_1(z)&=&\frac{1}{A}\frac{1}{z(\frac{\overline{A}}{A}-z)}\\
 b_2(z)&=&-\frac{1}{A}\frac{\sum\limits_{k=m}^Md_kz^{k+1}}
 {z(1-z)(\frac{\overline{A}}{A}-z)}\\
\end{eqnarray}
\begin{eqnarray}\label{th:2.5}
 e_i=-\frac{i\overline{A}+\overline{B}}{(i-1)(A+\overline{A})
 +1+B+\overline{B}},\ \  (i=1,2,\cdots,M)\\
 f_i=-\frac{(i-1)A+B}{i(A+\overline{A})+1+B+\overline{B}},
 \quad(i=1,2,\cdots,M-1)
\end{eqnarray}
\begin{eqnarray}\label{th:2.6}
 D=\left(\begin{array}{cccccccc}
 1  & 0         &         &        &         &            &      & \\
 1  &e_1        &         &        &         &            &      & \\
 f_1& 1         &e_2      &        &         &            &      & \\
 & \ddots    &\ddots   &\ddots  &         &            &      & \\
 &           &f_{i-1}  & 1      & e_i     &            &      & \\
 &           & \ddots  &\ddots  &\ddots   &            &      & \\
 &           &         & f_{M-2}& 1       & e_{M-1}    &      &\\
 &           &         &        & f_{M-1} & 1          & e_{M}g &\\
\end{array}\right).
\end{eqnarray}
\begin{eqnarray}\label{th:2.7}
 \overline{D}=\left(\begin{array}{cccccccc}
 \lim\limits_{\varepsilon\rightarrow0}\frac{-\int_\varepsilon^{h}
 b_2(s)e^{-\int_\varepsilon^sa(\theta)d\theta}ds}{1+\overline{B}
 \int_\varepsilon^hb_1(s)e^{-\int_\varepsilon^sa(\theta)d\theta}ds}\\
 d_0\\
 \vdots\\
 d_M\\
\end{array}\right).
\end{eqnarray}
\begin{eqnarray}\label{th:2.8}
 g=\left\{\begin{array}{ll}
 \int_0^1z^{M-1+\frac{1+\overline{B}}{\overline{A}}}(1-z)^{-\frac{1}
 {\overline{A}}}e^{\frac{B}{\overline{A}}\frac{1}{z}}dz,
 &\textrm{$A=0,\overline{A}\neq0$}\\
 \int_0^1z^{M-1+\frac{B}{A}}(1-z)^{\frac{1}{A}}e^{-\frac{\overline{B}}
 {A}z}dz, &\textrm{$\overline{A}=0,A\neq0$}\\
 \int_0^1z^{M-1+\frac{B}{A}}(1-z)^{\frac{\overline{B}-B}{A}}
 e^{\frac{1}{A}\frac{1}{z-1}}dz, &\textrm{$\overline{A}=A\neq0$}\\
 \int_0^1z^{M-1+\frac{B}{A}}(1-z)^{\frac{1}{A-\overline{A}}}
 \left|z-\frac{A}{\overline{A}}\right|^{\frac{\overline{B}}{\overline{A}}
 -\frac{1}{A-\overline{A}}-\frac{B}{A}}dz, &\textrm{else}\\
\end{array}\right.
\end{eqnarray}
in which $D_j$ is a matrix obtained by replacing the $j$th column
with $\overline{D}$ in the matrix $D$, and $|D_j|$ is the
determinant of $D_j$. Then,
\begin{eqnarray}\label{th:2.9}
 C=\frac{det D_{M+1}}{g\sum\limits_{i=1}^{M}e_i}
\end{eqnarray}
\end{theorem}
\bigbreak

\begin{theorem}\label{th:3}
When $F^+(k)=Ak+B, F^-(k)=\overline{A}k+\overline{B}$, one has:
\medskip

\indent $(I)$ \begin{eqnarray}\label{th:3.1} P(k)\geq0, \ \
\sum\limits_{k=0}^\infty P(k)=1
\end{eqnarray}

\indent $(II)$ When $ A>\overline{A}$, the network is scale-free
with the scaling exponent $1+\frac{1}{A-\overline{A}}$. However,
when $A\leq\overline{A}$, the network is not scale-free.
\end{theorem}

\section{Proofs of the Main Results}

\begin{lemma}\label{le:1}
If $d_k:=\lim\limits_{t\rightarrow\infty}d_{k,t}$ exists and
satisfies $\sum\limits_{k=0}^\infty d_k=1$, then the following
relations are satisfied for $P(k),k=0,1,2,\cdots\,$:
\begin{eqnarray}\label{th:1.1}
 P(k)=\left\{\begin{array}{ll}
 \frac{F^-(1)}{1+F^+(0)}P(k+1)+\frac{d_0}{1+F^+(0)}, &\textrm{$k=0$}\\
 \frac{F^+(k-1)}{1+F^+(k)+F^-(k)}P(k-1)+\frac{F^-(k+1)}{1+F^+(k)
 +F^-(k)}P(k+1)+\frac{d_{k}}{1+F^+(k)+F^-(k)}, &\textrm{$k>0$}\\
\end{array}\right.
\end{eqnarray}
\end{lemma}
\bigbreak

\indent{\bf Proof.}\ It follows from the Markovian properties that
\begin{eqnarray}
P(0,i,t+1)=P(0,i,t)[1-f_t^+(0)]+P(1,i,t)f_t^-(1)
\end{eqnarray}
Then, by the definitions of $P(k,t)$ and $P(0,i,i)=d_{0,i}$, one
obtains
\begin{eqnarray}
 P(0,t+1)=\frac{t}{t+1}P(0,t)[1-f_t^+(0)]+\frac{t}{t+1}P(1,t)f_t^-(1)
 +\frac{1}{t+1}d_{0,t+1}
\end{eqnarray}

\indent The above difference equation has the following solution:
\begin{eqnarray}
 P(0,t)&=&\frac{1}{t}\prod_{i=1}^{t-1}[1-f_i^+(0)]\nonumber\\
 &\times&\left\{P(0,1)+\sum_{l=1}^{t-1}\frac{P(1,l)lf_l^-(1)
 +d_{0,l+1}}{\prod_{j=1}^{l}[1-f_j^+(0)]}\right\}
\end{eqnarray}

\indent Let
\begin{eqnarray}
x_t&=&P(0,1)+\sum_{l=1}^{t-1}\frac{P(1,l)lf_l^-(1)+d_{0,l+1}}
{\prod_{j=1}^{l}[1-f_j^+(0)]}\\
y_t&=&t\prod_{i=1}^{t-1}[1-f_i^+(0)]^{-1}
\end{eqnarray}
Then, one can easily get
\begin{eqnarray}
 x_{t+1}-x_t&=&\frac{P(1,t)tf_t^-(1)
 +d_{0,t+1}}{\prod_{j=1}^{t}[1-f_j^+(0)]}\\
 y_{t+1}-y_t&=&[1+tf_t^+(0)]\prod_{i=1}^{t}[1-f_i^+(0)]^{-1}
\end{eqnarray}

\indent With the given condition, one has
\begin{eqnarray}
\frac{x_{t+1}-x_t}{y_{t+1}-y_t}=\frac{P(1,t)tf_t^-(1)
+d_{0,t+1}}{1+tf_t^+(0)}\rightarrow\frac{F^-(1)P(1)+d_0}{1+F^+(0)}
\end{eqnarray}

\indent With $P(0,t)=\frac{x_t}{y_t}$ and by the Stolz Theorem
$^{[6]}$, one obtains
\begin{eqnarray}
 P(0)=\frac{F^-(1)}{1+F^+(0)}P(1)+\frac{d_0}{1+F^+(0)}
\end{eqnarray}
When $k>0$, one has
\begin{eqnarray}
 P(k,i,t+1)&=&P(k,i,t)[1-f_t^+(k)-f_t^-(k)]
 +P(k+1,i,t)f_t^-(k+1)\nonumber\\
 &+&P(k-1,i,t)f_t^+(k-1)
\end{eqnarray}

\indent Similar to the above, one has
\begin{eqnarray}
 P(k)&=&\frac{F^+(k-1)}{1+F^+(k)+F^-(k)}P(k-1)
 +\frac{F^-(k+1)}{1+F^+(k)+F^-(k)}P(k+1)\nonumber\\
 &+&\frac{d_{k}}{1+F^+(k)+F^-(k)}
\end{eqnarray}

\indent Thus, the Lemma is proved.\quad $\Box$ \bigbreak

\begin{lemma}\label{le:2}
If there are constants $A,\ B,\ \overline{A}$ and $\overline{B}$,
such that $F^+(k)=Ak+B$ and $F^-(k)=\overline{A}k+\overline{B}$,
then
\begin{eqnarray}
 P(k)=\left\{\begin{array}{ll}
 \frac{\overline{A}+\overline{B}}{1+B}P(k+1)+\frac{d_0}{1+B},
 &\textrm{$k=0$}\\
 \frac{A(k-1)+B}{1+Ak+B+\overline{A}k+\overline{B}}P(k-1)
 +\frac{\overline{A}(k+1)+\overline{B}}{1+Ak+B+\overline{A}k
 +\overline{B}}P(k+1)+\frac{d_k}{1+Ak+B+\overline{A}k+\overline{B}},
 &\textrm{$k>0$}\\
\end{array}\right.
\end{eqnarray}
\end{lemma}
\bigbreak

\indent{\bf Proof.}\ It follows immediately from Lemma \ref{le:1}.
\quad $\Box$ \bigbreak

\indent Proof of Theorem \ref{th:1}: \medskip

\indent {\bf Proof.}\ The theorem follows easily from Lemmas
\ref{le:1} and \ref{le:2}.\quad $\Box$ \bigbreak

\begin{lemma}\label{le:3}
Suppose that $f_t^+(k)=a_tk+b_t+o(\frac{1}{t})$ and
$f_t^-(k)=\overline{a}_tk +\overline{b}_t+o(\frac{1}{t})$. Then,
$\lim\limits_{t\rightarrow\infty}tf_t^+(k)=Ak+B$ and
$\lim\limits_{t\rightarrow\infty}tf_t^-(k)=\overline{A}k+\overline{B}$
if and only if $\lim\limits_{t\rightarrow\infty}ta_t=A$,
$\lim\limits_{t\rightarrow\infty}tb_t=B$,
$\lim\limits_{t\rightarrow\infty}t\overline{a}_t=\overline{A}$, and
$\lim\limits_{t\rightarrow\infty}t\overline{b}_t=\overline{B}$.
\end{lemma}
\bigbreak

\begin{lemma}\label{le:4}
Suppose that $F^+(k)=Ak+B$, $F^-(k)=\overline{A}k+\overline{B}$, and
there are $0\leq m \leq M <\infty $ such that $d_k=0$ when $k< m$ or
$M> k$. Then,
\begin{eqnarray}\label{le:4.1}
 P(0)=\lim\limits_{\varepsilon\rightarrow0}\frac{-\int_\varepsilon^{h}b_2(s)
 e^{-\int_\varepsilon^sa(\theta)d\theta}ds}{1+\overline{B}\int_\varepsilon^h
 b_1(s)e^{-\int_\varepsilon^sa(\theta)d\theta}ds}
\end{eqnarray}
where $h,a(z),b_1(z),b_2(z)$ are given in Theorem \ref{th:2}.
\end{lemma}
\bigbreak

\indent{\bf Proof.}\ Let $F(z)=\sum\limits_{k=0}^{\infty} P(k)z^k$.
Then, one has $F(0)=P(0)$. With Eq. (\ref{th:1.2}) and the given
condition $d_k=0$, when $k< m$ and $M> k$, one obtains
\begin{eqnarray}\label{le:4.2}
 Az(1-z)\left(\frac{\overline{A}}{A}-z\right)F'(z)
 =-[Bz^2-(1+B+\overline{B})z+\overline{B}]F(z)
 +\overline{B}P(0)(1-z)-\sum\limits_{k=m}^Md_kz^{k+1}\nonumber\\
\end{eqnarray}

\indent Solving the above equation gives
\begin{eqnarray}\label{le:4.3}
 F(z)&=&F(\varepsilon)e^{\int_{\varepsilon}^za(\theta)d\theta}
 +\overline{B}P(0)e^{\int_{\varepsilon}^za(\theta)d\theta}
 \int_{\varepsilon}^zb_1(s)e^{-\int_{\varepsilon}^sa(\theta)d\theta}ds
 \nonumber\\
 &+&e^{\int_{\varepsilon}^za(\theta)d\theta}\int_{\varepsilon}^z
 b_2(s)e^{-\int_{\varepsilon}^sa(\theta)d\theta}ds
\end{eqnarray}
where $\varepsilon>0$ is small, and
\begin{eqnarray}\label{le:4.4}
 \frac{1}{e^{\int_\varepsilon^za(\theta)d\theta}}F(z)
 =F(\varepsilon)+\overline{B}P(0)\int_\varepsilon^zb_1(s)
 e^{-\int_\varepsilon^sa(\theta)d\theta}ds+\int_{\varepsilon}^z
 b_2(s)e^{-\int_{\varepsilon}^sa(\theta)d\theta}ds
\end{eqnarray}

\indent When $z\uparrow h$, the left hand of Eq. (\ref{le:4.4}) is
0, so
\begin{eqnarray}
 F(\varepsilon)+\overline{B}P(0)\int_\varepsilon^hb_1(s)
 e^{-\int_\varepsilon^sa(\theta)d\theta}ds+\int_{\varepsilon}^h
 b_2(s)e^{-\int_{\varepsilon}^sa(\theta)d\theta}ds=0
\end{eqnarray}

\indent With
$P(0)=\lim\limits_{\varepsilon\downarrow0}F(\varepsilon)$, and by
letting $\varepsilon\downarrow0$, one obtains
\begin{eqnarray}
 P(0)=\lim\limits_{\varepsilon\rightarrow0}\frac{-\int_\varepsilon^{h}
 b_2(s)e^{-\int_\varepsilon^sa(\theta)d\theta}ds}{1+\overline{B}
 \int_\varepsilon^hb_1(s)e^{-\int_\varepsilon^sa(\theta)d\theta}ds}
\end{eqnarray}

\indent Since $P(0)$ is uniquely determined, the solution of Eq.
(\ref{th:1.2}), i.e., $P(k)$, $k=0,1,2,\cdots\,$, is unique. \quad
$\Box$ \bigbreak

\begin{lemma}\label{le:5}
If $A$ and $\overline{A}$ are not both 0, then when $k\geq M$, Eq
(\ref{th:1.2}) has the following solutions:
\begin{eqnarray}\label{le:5.1}
 P(k)=\left\{\begin{array}{ll}
 C\int_0^1z^{k-1+\frac{1+\overline{B}}{\overline{A}}}(1-z)^{-\frac{1}
 {\overline{A}}}e^{\frac{B}{\overline{A}}\frac{1}{z}}dz,
 &\textrm{$A=0,\overline{A}\neq0$}\\
 C\int_0^1z^{k-1+\frac{B}{A}}(1-z)^{\frac{1}{A}}e^{-\frac{\overline{B}}{A}z}dz,
 &\textrm{$\overline{A}=0,A\neq0$}\\
 C\int_0^1z^{k-1+\frac{B}{A}}(1-z)^{\frac{\overline{B}-B}{A}}
 e^{\frac{1}{A}\frac{1}{z-1}}dz, &\textrm{$\overline{A}=A\neq0$}\\
 C\int_0^1z^{k-1+\frac{B}{A}}(1-z)^{\frac{1}{A-\overline{A}}}
 \left|z-\frac{A}{\overline{A}}\right|^{\frac{\overline{B}}{\overline{A}}
 -\frac{1}{A-\overline{A}}-\frac{B}{A}}dz, &\textrm{else}\\
\end{array}\right.
\end{eqnarray}
where $C$ is a constant.
\end{lemma}
\bigbreak

\indent {\bf Proof.}\ It is easily to see that (\ref{le:5.1})
satisfies Eq. (\ref{th:1.2}). \quad  $\Box$ \bigbreak

\indent Proof of Theorem \ref{th:2}: \bigbreak

\indent {\bf Proof.}\ From Lemma \ref{le:5} and Eqs. (\ref{th:1.2})
and (\ref{le:5.1}), one has
\begin{eqnarray}
\left\{\begin{array}{ll}
 P(0)=\lim\limits_{\varepsilon\rightarrow0}\frac{-\int_\varepsilon^{h}
 b_2(s)e^{-\int_\varepsilon^sa(\theta)d\theta}ds}{1+\overline{B}
 \int_\varepsilon^hb_1(s)e^{-\int_\varepsilon^sa(\theta)d\theta}ds}\\
 P(0)=\frac{\overline{A}+\overline{B}}{1+B}P(1)+\frac{d_0}{1+B}\\
 P(1)=\frac{B}{\overline{A}+A+1+B+\overline{B}}P(0)
 +\frac{2\overline{A}+\overline{B}}{\overline{A}+A+1+B+
 \overline{B}}P(2)+\frac{d_1}{\overline{A}+A+1+B+\overline{B}}\\
 \vdots\\
 P(m)=\frac{A(m-1)+B}{(A+\overline{A})m+1+B+\overline{B}}P(m-1)
 +\frac{\overline{A}(m+1)+\overline{B}}{(A+\overline{A})m+1+B+
 \overline{B}}P(m+1)+\frac{d_m}{(A+\overline{A})m+1+B+\overline{B}}\\
 \vdots\\
 P(M-1)=\frac{A(M-2)+B}{(A+\overline{A})(M-1)+1+B+\overline{B}}P(M-2)
 +\frac{\overline{A}(M+1)+\overline{B}}{(A+\overline{A})M+1+B+
 \overline{B}}Cg+\frac{d_M}{(A+\overline{A})M+1+B+\overline{B}}
\end{array}\right.
\end{eqnarray}

\indent This is a system of equations with $M+1$ unknown variables,
where $g$ is given by Eq. (\ref{th:2.8}). Solving this system of
equations, one obtains (\ref{th:2.1}), and (\ref{th:2.2}) is
obtained by substituting $C$ into (\ref{le:5.1}). \quad
 $\Box$ \bigbreak

\begin{lemma}\label{le:6}
When $A\leq\overline{A}$, $P(k)$ is not power-law.
\end{lemma}
\bigbreak

\indent {\bf Proof.}\ When $k>m$, Eq. (\ref{th:1.2}) can be
rewritten as
\begin{eqnarray}\label{le:6.1}
 [(A+\overline{A})k+1+B+\overline{B}]P(k)=[A(k-1)+B]P(k-1)
 +[\overline{A}(k+1)+\overline{B}]P(k+1)
\end{eqnarray}

\indent Suppose that $P(k)$ is power-law. Then, one has
$P(k)=Ck^{-\gamma}[1+o_k(1)]$, where $\gamma>1$ is the scaling
exponent, $C$ is a constant, and $o_k(1)$ is an infinitesimal with
respect to $k$. It follows that
\begin{eqnarray}\label{le:6.2}
 & &[1+(A+\overline{A})k+B+\overline{B}]k^{-\gamma}[1+o_k(1)]
 =[A(k-1)+B](k-1)^{-\gamma}[1+o_{k-1}(1)]\nonumber\\
 & &+[\overline{A}(k+1)+\overline{B}](k+1)^{-\gamma}[1+o_{k+1}(1)]
\end{eqnarray}
that is,
\begin{eqnarray}\label{le:6.3}
 & &\left(A+\overline{A}+\frac{1+B+\overline{B}}{k}\right)
 \left(1-\frac{1}{k}\right)^\gamma\left(1+\frac{1}{k}\right)^\gamma
 -\left(A+\frac{B-A}{k}\right)\left(1+\frac{1}{k}\right)^\gamma\nonumber\\
 & &-\left(\overline{A}+\frac{\overline{A}+\overline{B}}{k}\right)
 \left(1-\frac{1}{k}\right)^\gamma
 =-\left(A+\overline{A}+\frac{1+B+\overline{B}}{k}\right)
 \left(1-\frac{1}{k}\right)^\gamma
 \left(1+\frac{1}{k}\right)^\gamma o_k(1)\nonumber\\
 & &+\left(A+\frac{B-A}{k}\right)\left(1+\frac{1}{k}\right)^\gamma
 o_{k-1}(1)+\left(\overline{A}+\frac{\overline{A}+\overline{B}}{k}\right)
 \left(1-\frac{1}{k}\right)^\gamma o_{k+1}(1)
\end{eqnarray}
The first term on the left of the above expansion is
$[(1+A-\overline{A})-(A-\overline{A})\gamma]\frac{1}{k}$, the first
term of the right is
$[-(A+\overline{A})o_k(1)+Ao_{k-1}(1)+\overline{A}o_{k+1}(1)]$.
These two terms must be equal after neglecting the high-order
infinitesimals; that is,
\begin{eqnarray}\label{le:6.4}
 [(1+A-\overline{A})-(A-\overline{A})\gamma]\frac{1}{k}
 =-(A+\overline{A})o_k(1)+Ao_{k-1}(1)+\overline{A}o_{k+1}(1)
\end{eqnarray}
Thus, summing over $k$, one obtained
\begin{eqnarray}\label{le:6.5}
 [(1+A+\overline{A})-(A-\overline{A})\gamma]\sum_{k=k_0}^\infty\frac{1}{k}
 =-\overline{A}o_{k_{0}}(1)+Ao_{k_0-1}(1)
\end{eqnarray}
To this end, one has $(1+A-\overline{A})-(A-\overline{A})\gamma=0$
since $o_k(1)$ is a infinitesimal, so that
$\gamma=1+\frac{1}{A-\overline{A}}$. And, since $\gamma>1$, one has
$A>\overline{A}$. From the assumption, the proof is compete. \quad
$\Box$ \bigbreak

\begin{lemma}\label{le:7}
\indent $(I)$ When $A>\overline{A}=0$,
\begin{eqnarray}\label{le:7.1}
 P(k)=C\int_0^1z^{k-1+\frac{B}{A}}(1-z)^{\frac{1}{A}}
 e^{-\frac{\overline{B}}{A}z}dz
 \sim e^{-\frac{\overline{B}}{A}}k^{-(1+\frac{1}{A})}
\end{eqnarray}

\indent$(II)$ When $A>\overline{A}>0$,
\begin{eqnarray}\label{le:7.2}
 P(k)&=&C\int_0^1z^{k-1+\frac{B}{A}}(1-z)^{\frac{1}{A-\overline{A}}}
 \left(\frac{A}{\overline{A}}-z\right)^{\frac{\overline{B}}{\overline{A}}
 +\frac{1}{A-\overline{A}}-\frac{B}{A}}dz\nonumber\\
 &\sim&k^{-(1+\frac{1}{A-\overline{A}})}
\end{eqnarray}
\end{lemma}
\bigbreak

\indent {\bf Proof.}\ (I) When $A>\overline{A}=0$, one has
$\lim\limits_{k\rightarrow\infty}k^\gamma\frac{\Gamma(k+k_0)}
{\Gamma(k+k_0+\gamma)}=1$, where $\gamma,k_0$ are non-negative real
numbers, i.e., there is a number $K$ satisfying
$k^\gamma\frac{\Gamma(k)}{\Gamma(k+\gamma)}<1+\varepsilon<2$ when
$k>K$, where $\varepsilon>0$ can be arbitrary.

\indent When $k>K$, one has
\begin{eqnarray}
 \left|\,\frac{P(k)}{k^{-(1+\frac{1}{A})}}\right|&=&\left|k^{1+\frac{1}{A}}P(k)\,\right|
 =\left|\,k^{1+\frac{1}{A}}C\int_0^1z^{k-1+\frac{B}{A}}(1-z)^{\frac{1}{A}}
 e^{-\frac{\overline{B}}{A}z}dz\,\right|\nonumber\\
 &=&\left|\,Ck^{1+\frac{1}{A}}\int_0^1z^{k-1+\frac{B}{A}}(1-z)^{\frac{1}{A}}
 \sum_{s=0}^\infty\frac{(-\frac{\overline{B}}{A}z)^s}{s!}dz\,\right|\nonumber\\
 &=&\left|\,C\sum_{s=0}^\infty\frac{1}{s!}\left(-\frac{\overline{B}}{A}\right)^s
 k^{1+\frac{1}{A}}\int_0^1z^{k+s+\frac{B}{A}-1}(1-z)^{\frac{1}{A}}dz\,\right|\nonumber\\
 &=&\left|\,C\sum_{s=0}^\infty\frac{1}{s!}\left(-\frac{\overline{B}}{A}\right)^s
 k^{1+\frac{1}{A}}\frac{\Gamma(k+s+\frac{B}{A})\Gamma(1+\frac{1}{A})}
 {\Gamma(k+s+\frac{B}{A}+1+\frac{1}{A})}\,\right|\nonumber\\
 &\leq&\left|\,C\sum_{s=0}^\infty\frac{1}{s!}\left(\frac{\overline{B}}{A}\right)^s\,\right|
 k^{1+\frac{1}{A}}\frac{\Gamma(k+s+\frac{B}{A})\Gamma(1+\frac{1}{A})}
 {\Gamma(k+s+\frac{B}{A}+1+\frac{1}{A})}\nonumber\\
 &\leq&\left|\,C\sum_{s=0}^\infty\frac{1}{s!}\left(\frac{\overline{B}}{A}\right)^s\,\right|
 \left(k+s+\frac{B}{A}\right)^{1+\frac{1}{A}}\frac{\Gamma(k+s+\frac{B}{A})
 \Gamma(1+\frac{1}{A})}{\Gamma(k+s+\frac{B}{A}+1+\frac{1}{A})}\nonumber\\
 &\leq&\left|\,C\sum_{s=0}^\infty\frac{1}{s!}\left(\frac{\overline{B}}{A}\right)^s\,\right|
 2\Gamma(1+\frac{1}{A})+1\nonumber\\
 &=&2|C|\,\Gamma\left(1+\frac{1}{A}\right)e^{\frac{\overline{B}}{A}}<+\infty
\end{eqnarray}

\indent Thus, one obtains
\begin{eqnarray}
 \lim_{k\rightarrow\infty}\frac{P(k)}{k^{-(1+\frac{1}{A})}}
 &=&\lim_{k\rightarrow\infty}C\sum_{s=0}^\infty\frac{1}{s!}
 \left(-\frac{\overline{B}}{A}\right)^s
 k^{1+\frac{1}{A}}\frac{\Gamma(k+s+\frac{B}{A})\Gamma(1+\frac{1}{A})}
 {\Gamma(k+s+\frac{B}{A}+1+\frac{1}{A})}\nonumber\\
 &=&C\sum_{s=0}^\infty\frac{1}{s!}\left(-\frac{\overline{B}}{A}\right)^s
 \left(\lim_{k\rightarrow\infty}k^{1+\frac{1}{A}}\frac{\Gamma(k+s+\frac{B}{A})
 \Gamma(1+\frac{1}{A})}{\Gamma(k+s+\frac{B}{A}+1+\frac{1}{A})}\right)\nonumber\\
 &=&C\sum_{s=0}^\infty\frac{1}{s!}\left(-\frac{\overline{B}}{A}\right)^s
 \Gamma\left(1+\frac{1}{A}\right)\nonumber\\
 &=&C\Gamma\left(1+\frac{1}{A}\right)e^{-\frac{\overline{B}}{A}}
\end{eqnarray}

\indent Consequently, one has
\begin{eqnarray}
 P(k)&=&C\int_0^1z^{k-1+\frac{B}{A}}(1-z)^{\frac{1}{A}}
 e^{-\frac{\overline{B}}{A}z}dz\nonumber\\
 &\sim&C\Gamma\left(1+\frac{1}{A}\right)e^{-\frac{\overline{B}}{A}}k^{-(1+\frac{1}{A})}
\end{eqnarray}

(II) When $A>\overline{A}>0$, one has
\begin{eqnarray}
 \left|\,\frac{P(k)}{k^{-(1+\frac{1}{A-\overline{A}})}}\,\right|
 &=&\left|\,k^{1+\frac{1}{A-\overline{A}}}P(k)\right|=\left|\,
 k^{1+\frac{1}{A-\overline{A}}}C\int_0^1z^{k-1+\frac{B}{A}}
 (1-z)^{\frac{1}{A-\overline{A}}}\left(\frac{A}{\overline{A}}
 -z\right)^{\frac{\overline{B}}{\overline{A}}-\frac{1}{A-\overline{A}}
 -\frac{B}{A}}dz\,\right|\nonumber\\
 &=&\left|\,C\left(\frac{A}{\overline{A}}\right)^{\frac{\overline{B}}
 {\overline{A}}-\frac{1}{A-\overline{A}}-\frac{B}{A}}
 k^{1+\frac{1}{A-\overline{A}}}
 \int_0^1z^{k-1+\frac{B}{A}}(1-z)^{\frac{1}{A-\overline{A}}}
 \left(1-\frac{\overline{A}}{A}z\right)^{\frac{\overline{B}}{\overline{A}}
 -\frac{1}{A-\overline{A}}-\frac{B}{A}}dz\,\right|\nonumber\\
 &=&\left|\,C\left(\frac{A}{\overline{A}}\right)^{\frac{\overline{B}}
 {\overline{A}}-\frac{1}{A-\overline{A}}-\frac{B}{A}}k^{1+\frac{1}
 {A-\overline{A}}}\int_0^1z^{k-1+\frac{B}{A}}(1-z)^{\frac{1}{A-\overline{A}}}
 \sum_{s=0}^\infty H_s\left(\frac{\overline{A}}{A}z\right)^sdz\,\right|\nonumber\\
 &=&\left|\,C\left(\frac{A}{\overline{A}}\right)^{\frac{\overline{B}}{\overline{A}}
 -\frac{1}{A-\overline{A}}-\frac{B}{A}}\sum_{s=0}^\infty
 H_s\left(\frac{\overline{A}}{A}\right)^sk^{1+\frac{1}{A-\overline{A}}}\int_0^1
 z^{k+s-1+\frac{B}{A}}(1-z)^{\frac{1}{A-\overline{A}}}dz\,\right|\nonumber\\
 &\leq &\left|\,C\left(\frac{A}{\overline{A}}\right)^{\frac{\overline{B}}
 {\overline{A}}-\frac{1}{A-\overline{A}}-\frac{B}{A}}\sum_{s=0}^\infty
 H_s\left(\frac{\overline{A}}{A}\right)^s\right|k^{1+\frac{1}{A-\overline{A}}}
 \frac{\Gamma(k+s+\frac{B}{A})\Gamma(1+\frac{1}{A-\overline{A}})}
 {\Gamma(k+s+\frac{B}{A}+1+\frac{1}{A-\overline{A}})}\nonumber\\
 &=&\left|\,C\left(\frac{A}{\overline{A}}\right)^{\frac{\overline{B}}
 {\overline{A}}-\frac{1}{A-\overline{A}}-\frac{B}{A}}\sum_{s=0}^\infty
 H_s\left(\frac{\overline{A}}{A}\right)^s\right|2\Gamma(1+\frac{1}
 {A-\overline{A}})\nonumber\\
 &=&2|C|\,\Gamma\left(1+\frac{1}{A-\overline{A}}\right)
 \left(\frac{A}{\overline{A}}+1\right)^{\frac{\overline{B}}{\overline{A}}
 -\frac{1}{A-\overline{A}}-\frac{B}{A}}<+\infty
\end{eqnarray}
where $H_s$ is the coefficient of $(\frac{\overline{A}}{A}z)^s$ in
the expansion of $(1-\frac{\overline{A}}{A}z)^{\frac{\overline{B}}
{\overline{A}}-\frac{1}{A-\overline{A}}-\frac{B}{A}}$.

\indent It follows that
\begin{eqnarray}
 & &\lim_{k\rightarrow\infty}\frac{P(k)}{k^{-(1+\frac{1}{A-\overline{A}})}}
 =\lim_{k\rightarrow\infty}k^{1+\frac{1}{A-\overline{A}}}C\int_0^1
 z^{k-1+\frac{B}{A}}(1-z)^{\frac{1}{A-\overline{A}}}\left(\frac{A}
 {\overline{A}}-z\right)^{\frac{\overline{B}}{\overline{A}}
 -\frac{1}{A-\overline{A}}-\frac{B}{A}}dz\nonumber\\
 & &=\lim_{k\rightarrow\infty}C\left(\frac{A}{\overline{A}}
 \right)^{\frac{\overline{B}}
 {\overline{A}}-\frac{1}{A-\overline{A}}-\frac{B}{A}}k^{1+\frac{1}
 {A-\overline{A}}}\int_0^1z^{k-1+\frac{B}{A}}(1-z)^{\frac{1}{A-\overline{A}}}
 \sum_{s=0}^\infty H_s\left(\frac{\overline{A}}{A}z\right)^sdz\nonumber\\
 & &=\lim_{k\rightarrow\infty}C\left(\frac{A}{\overline{A}}
 \right)^{\frac{\overline{B}}
 {\overline{A}}-\frac{1}{A-\overline{A}}-\frac{B}{A}}\sum_{s=0}^\infty
 H_s\left(\frac{\overline{A}}{A}\right)^sk^{1+\frac{1}{A-\overline{A}}}
 \frac{\Gamma(k+s+\frac{B}{A})\Gamma(1+\frac{1}{A-\overline{A}})}
 {\Gamma(k+s+\frac{B}{A}+1+\frac{1}{A-\overline{A}})}\nonumber\\
 & &=C\left(\frac{A}{\overline{A}}\right)^{\frac{\overline{B}}{\overline{A}}
 -\frac{1}{A-\overline{A}}-\frac{B}{A}}\sum_{s=0}^\infty
 H_s\left(\frac{\overline{A}}{A}\right)^s\left(\lim_{k\rightarrow\infty}
 k^{1+\frac{1}{A-\overline{A}}}\frac{\Gamma(k+s+\frac{B}{A})
 \Gamma(1+\frac{1}{A-\overline{A}})}{\Gamma(k+s+\frac{B}{A}+1+
 \frac{1}{A-\overline{A}})}\right)\nonumber\\
 & &=C\left(\frac{A}{\overline{A}}\right)^{\frac{\overline{B}}{\overline{A}}
 -\frac{1}{A-\overline{A}}-\frac{B}{A}}\sum_{s=0}^\infty
 H_s\left(\frac{\overline{A}}{A}\right)^s
 \Gamma\left(1+\frac{1}{A-\overline{A}}\right)\nonumber\\
 & &=C\Gamma\left(1+\frac{1}{A-\overline{A}}\right)\left(\frac{A}
 {\overline{A}}-1\right)^{\frac{\overline{B}}{\overline{A}}
 +\frac{1}{A-\overline{A}}-\frac{B}{A}}
\end{eqnarray}

\indent Thus,
\begin{eqnarray}
 P(k)&=&C\int_0^1z^{k-1+\frac{B}{A}}(1-z)^{\frac{1}{A-\overline{A}}}
 \left(\frac{A}{\overline{A}}-z\right)^{\frac{\overline{B}}{\overline{A}}
 -\frac{1}{A-\overline{A}}-\frac{B}{A}}dz\nonumber\\
 &\sim&C\Gamma\left(1+\frac{1}{A-\overline{A}}\right)
 \left(\frac{A}{\overline{A}}-1\right)^{\frac{\overline{B}}{\overline{A}}
 -\frac{1}{A-\overline{A}}-\frac{B}{A}}k^{-(1+\frac{1}{A-\overline{A}})}
\end{eqnarray}

\indent The Lemma is proved. \quad $\Box$ \bigbreak

\indent Proof of Theorem \ref{th:3}: \bigbreak

\indent {\bf Proof.} Summing $k$ from $0$ to $2^n$ in Eq.
(\ref{th:1.2}), where $n$ is an integer, gives
\begin{eqnarray}
 \sum\limits_{k=0}^{2^n}P(k)+A2^nP(2^n)+BP(2^n)
 =\overline{A}(2^n+1)P(2^n+1)+\overline{B}P(2^n+1)
 +\sum\limits_{k=0}^{2^n}d_k
\end{eqnarray}
along with $P(k)\geq 0$, $\sum\limits_{k=0}^{\infty}P(k)\leq1$ and
Eq. (\ref{le:5.1}), one has $P(k)\downarrow0$ when $k>M$ and
$k\uparrow \infty$, and $\sum_{k=0}2^kP(2^k)<\infty$. Moreover, one
has $(2^k+s)P(2^k+s)\rightarrow0$ when $k\rightarrow\infty$, where
$s$ is an integer. Letting $n\rightarrow\infty$ in Eq. (54) yields
$\sum\limits_{k=0}^{\infty}P(k)=1$.

\indent From Lemma \ref{le:7}, one can see that $P(k)$ is power-law
with scaling exponent $1+\frac{1}{A-\overline{A}}$ when
$A>\overline{A}$. From Lemma \ref{le:6}, one can see that the
network is not scale-free when $A\leq\overline{A}$. \quad $\Box$

\section{Examples}

\begin{example}\label{ex:1}

\indent Start with a small number $m_0$ of nodes, which together
have a total degree $N_0$.

\indent At each time step, perform the following two operations
independently.

\indent (i) Add a new node with $m$ $(1<m\leq m_0)$ edges that link
the new node to $m$ different nodes already present in the network.
And, the preferential probability is similar to that in the BA
model, i.e., the probability that the new node is connected to an
old node $i$ depends on the connectivity (degree) $k_i$ of that
node; that is,
\begin{eqnarray}\label{ex:1.1}
\Pi(k_i)=\frac{k_i}{\sum_jk_j}
\end{eqnarray}

\indent (ii) Delete an old edge. In so doing, select a node $i$ with
probability $\Pi'(k_i)$ given by Eq. (\ref{ex:1.1}), and select a
node $j$ at random in the domain of $i$; then, remove the edge
$l_{ij}$. After $t$ steps, the model becomes a random network with
$t+m_0\approx t$ nodes with the total degree $2(m-1)t+N_0\approx
2(m-1)t$.

\indent From (i), one can see that the probability
$\Pi_t^+(k_i{(t)})$ for node $i$ to increase its degree $k_i(t)$ by
one is
\begin{eqnarray}\label{ex:1.2}
 \Pi_t^+(k_i(t))=m\Pi(k_i(t))=m\frac{k_i(t)}{\sum_j^tk_j(t)}
 =m\frac{k_i(t)}{2(m-1)t}
\end{eqnarray}

\indent From (ii), one obtains the probability $\Pi_t^-(k_i{(t)})$
for node $i$ to decrease its degree $k_i(t)$ by one, which is
\begin{eqnarray}\label{ex:1.3}
 \Pi_t^-(k_i(t))&=&\Pi'(k_i(t))+\sum_{j\in O_i}\Pi'(k_j(t))
 \frac{1}{k_j(t)}\nonumber\\
 &=&\frac{k_i(t)}{2(m-1)t}+\sum_{j\in O_i}\frac{1}{2(m-1)t}\nonumber\\
 &=&2\frac{k_i(t)}{2(m-1)t}=\frac{k_i(t)}{(m-1)t}
\end{eqnarray}
where $\Pi'(k_i(t))$ is the probability of node $i$ to be selected
preferentially, and $\sum_{j\in O_i}\Pi'(k_j(t))\frac{1}{k_j(t)}$ is
the probability of node $i$ to be selected randomly.

\indent Consider Eq. (\ref{ex:1.2}) and Eq. (\ref{ex:1.3}). One has
the following transition probability:
\begin{eqnarray}
 P\{k_i(t+1)=l|k_i(t)=k\}=\left\{\begin{array}{ll}
 \Pi_t^+(k)[1-\Pi_t^-(k)]
 =\frac{mk}{2(m-1)t}-\frac{mk^2}{2(m-1)^2t^2}, &\textrm{$l=k+1$}\\
 \Pi_t^-(k)[1-\Pi_t^+(k)]
 =\frac{k}{(m-1)t}-\frac{mk^2}{2(m-1)^2t^2}, &\textrm{$l=k-1$}\\
 1-\Pi_t^+(k)-\Pi_t^+(k)
 =1-\frac{(m+2)k}{2(m-1)t}+\frac{mk^2}{(m-1)^2t^2}, &\textrm{$l=k$}\\
 0. &\textrm{otherwise}\\
\end{array}\right.
\end{eqnarray}
and $A=\frac{m}{2(m-1)}$, $\overline{A}=\frac{1}{m-1}$,
$B=\overline{B}=0$, $d_m=1$.

\indent When $m=2$ ($A=\overline{A}$), this network is not
scale-free, and
\begin{eqnarray}\label{ex:1.4}
 P(0)=e\int_0^1\frac{s^2}{(1-s)^2}e^{-\frac{1}{1-s}}ds
\end{eqnarray}

\indent When $m>2$($A>\overline{A})$, the network is scale-free and
\begin{eqnarray}\label{ex:1.5}
 P(0)=\int_0^{\frac{2}{m}}\frac{z^{m}}{(z-1)(\frac{m}{2(m-1)}z
 -\frac{1}{m-1})}e^{-\int_0^z\frac{1}{(s-1)(\frac{m}{2(m-1)}s
 -\frac{1}{m-1})}ds}dz\\
 P(k)=C\int_0^1z^{k-1}(1-z)^{\frac{2(m-1)}{m-2}}
 (\frac{m}{2}-z)^{-\frac{2(m-1)}{m-2}}dz \quad (k\geq m)
\end{eqnarray}

\indent One can easily obtain $P(m)$ from $P(0)$. Further, one can
obtain $C$. It is clear that $\sum\limits_{k=0}^\infty P(k)=1$ is an
distribution according to Theorem \ref{th:3}, and that $P(k)$ is
power-law with scaling exponent $3+\frac{2}{m-2}$ according to Lemma
\ref{le:7}, i.e.,
\begin{eqnarray}
 P(k)\thicksim
 C\Gamma(3+\frac{2}{m-2})(\frac{m}{2}-1)^{-(2+\frac{2}{m-2})}
 k^{-(1+\frac{2(m-1)}{m-2})}
\end{eqnarray}

\indent For instance, when $m=3$, one has
$A=\frac{3}{4},\overline{A}=\frac{1}{2}$, $B=\overline{B}=0, d_3=1$
and
\begin{eqnarray}
P(0)=47-\frac{171}{4}ln3
\end{eqnarray}

\indent It follows from Eq. (\ref{th:1.2}) that
\begin{eqnarray}
 P(1)&=&2P(0)=94-\frac{171}{2}ln3\nonumber\\
 P(2)&=&\frac{9}{2}P(0)=\frac{423}{2}-\frac{1539}{8}ln3\nonumber\\
 P(3)&=&\frac{19}{2}P(0)=\frac{19}{2}(47-\frac{171}{4}ln3)
\end{eqnarray}

\indent When $k\geq 3$, $P(k)$ has the following form:
\begin{eqnarray}
 P(k)=C\int_0^1z^{k-1}(1-z)^{4}(\frac{3}{2}-z)^{-4}dz
\end{eqnarray}
and
$C=\frac{\frac{19}{2}(47-\frac{171}{4}ln3)}{\int_0^1z^{2}(1-z)^{4}
(\frac{3}{2}-z)^{-4}dz}=\frac{171}{4}$. Furthermore,
$C\Gamma(1+\frac{1}{A-\overline{A}})(\frac{A}
{\overline{A}}-1)^{-\frac{1}{A-\overline{A}}}
=\frac{171}{4}\Gamma(5)(\frac{1}{2})^{-4}=16416$.

\indent Therefore, the degree distribution is
\begin{eqnarray}
 P(k)=\left\{\begin{array}{llll}
 47-\frac{171}{4}ln3, &\textrm{$k=0$}\\
 94-\frac{171}{2}ln3, &\textrm{$k=1$}\\
 \frac{423}{2}-\frac{1539}{8}ln3, &\textrm{$k=2$}\\
 \frac{171}{4}\int_0^1z^{k-1}(1-z)^{4}(\frac{3}{2}-z)^{-4}dz
 \thicksim 16416k^{-5},&\textrm{$k>2$}\\
\end{array}\right.
\end{eqnarray}
\end{example}
\bigbreak

\begin{example}\label{ex:2}

\indent Start with a small number $m_0$ of nodes, which together
have a total degree $N_0$.

\indent At each time step, add a new node with $m$ $(1<m\leq m_0)$
edges that link the new node to $m$ different nodes already present
in the network. The probability that the new node is connected to
$m$ old nodes is the group preferential attachment$^{[7]}$, i.e.,
the probability for an old node $i$ to receive one edge is
\begin{eqnarray}\label{ex:2.1}
 \Pi^+(k_i(t))=\frac{m_0+t-m}{m_0+t-1}\frac{k_i(t)}
 {\sum_jk_j(t)}+\frac{m-1}{m_0+t-1}
\end{eqnarray}
At the same time, remove an old edge. To do so, select a node $i$
with probability $\frac{k_i}{\sum_jk_j}$, and select a node $j$ at
random in the domain of $i$; then, remove the edge $l_{ij}$, i.e.,
the probability for the old node $i$ to remove one edge is
\begin{eqnarray}\label{ex:2.2}
 \Pi_t^-(k_i(t))&=&\frac{k_i(t)}{\sum_lk_l(t)}
 +\sum_{j\in O_i}\frac{k_j(t)}{\sum_lk_l(t)}\frac{1}{k_j(t)}\nonumber\\
 &=&\frac{k_i(t)}{2(m-1)t}+\sum_{j\in O_i}\frac{1}{2(m-1)t}\nonumber\\
 &=&2\frac{k_i(t)}{2(m-1)t}=\frac{k_i(t)}{(m-1)t}.
\end{eqnarray}

\indent After $t$ steps, the model becomes a random network with
$t+m_0\approx t$ nodes having the total degree $2(m-1)t+N_0\approx
2(m-1)t$.

\indent The probability for a node with degree $k$ to increase its
degree by one or to decrease its degree by one, denoted by
$f_t^+(k)$ or $f_t^-(k)$ respectively, is given by
\begin{eqnarray}\label{ex:2.2}
 & &f_t^+(k)=\left(\frac{t-m}{t-1}\frac{k}{2(m-1)t}
 +\frac{m-1}{t-1}\right)\left(1-\frac{k}{(m-1)t}\right)\\
 & &f_t^-(k)=\frac{k}{(m-1)t}\left(1-\frac{t-m}{t-1}
 \frac{k}{2(m-1)t}-\frac{m-1}{t-1}\right)
\end{eqnarray}

\indent one thus has $A=\frac{1}{2(m-1)}$, $B=m-1$,
$\overline{A}=\frac{1}{m-1}$, $\overline{B}=0$, $B=\overline{B}=0$,
and $d_m=1$.

\indent It follows that $A<\overline{A}$ when $m>1$, and
\begin{eqnarray}
 P(0)=2(m-1)\int_0^1s^m(1-s)^{2m-3}(2-s)^{2(m-1)(m-2)-1}ds
\end{eqnarray}

\indent One can see that this network is not scale-free by Lemma
\ref{le:6}.
\end{example}
\bigbreak

\begin{example}\label{ex:3}

\indent This model is a revised model proposed by Albert et
al.$^{[8]}$.

\indent Start with $m_0$ isolated nodes. At each time step, add a
new node and perform one of the following three operations:

\indent (i) With probability $p$, add $m$ $(m\leq m_0)$ new edges:
In so doing, randomly select one node as the starting point of the
new edge, and the other end of the edge is selected with probability
\begin{eqnarray}\label{ex:3.1}
 \Pi(k_i)=\frac{k_i+1}{\sum_j (k_j+1)}
\end{eqnarray}
Taking into account the fact that new edges preferentially point to
popular nodes with large numbers of connections. This process is
repeated $m$ times.

\indent (ii) With probability $q$, rewire $m$ edges: In so doing,
randomly select a node $i$ and an edge $l_{ij}$ connected to it.
Then, remove this edge and replace it with a new edge $l_{ij'}$ that
connects to $i$ with node $j'$ chosen, with probability $\Pi(k_j')$
given by (\ref{ex:3.1}). This process is repeated $m$ times.

\indent (iii) With probability $1-p-q$, the new node with $m$ new
edges are connected to $i$ nodes already present in the network,
with probability $\Pi(k_i)$.

\indent In this model, one has $d_0=p+q,d_m=1-p-q$ and the
probability $f_t^+(k)$ for a node with degree $k$ to increase its
degree by one is
\begin{eqnarray}
 f_t^+(k)&=&pm\left(\frac{1}{N}+\frac{k+1}{\sum_j(k_j+1)}\right)
 +qm\left[(1-\frac{1}{N})\frac{k+1}{\sum_j(k_j+1)}\right]
 +(1-p-q)\frac{k+1}{\sum_j(k_j+1)}\nonumber\\
 &=&m\frac{k+1}{\sum_j(k_j+1)}+pm\frac{1}{N}
 +qm\frac{k+1}{\sum_j(k_j+1)}\frac{1}{N}\nonumber\\
 &=&m\frac{k+1}{N+\sum_ik_i(i)}+pm\frac{1}{N}
 +qm\frac{k+1}{N+\sum_ik_i(i)}\frac{1}{N}\nonumber\\
 &=&m\frac{1}{t}\frac{k+1}{1+(1-q)2m+o(1)}
 +pm\frac{1}{t}+qm\frac{1}{t^2}\frac{k+1}{1+(1-q)2m+o(1)}\nonumber\\
\end{eqnarray}

\indent One thus obtains $A=\frac{m}{(1-q)2m+1}$,
$B=\frac{m}{(1-q)2m+1}+pm$.

\indent The probability $f_t^-(k)$ for a node with degree $k$ to
decrease its degree by one is
\begin{eqnarray}
 f_t^-(k)&=&qm\frac{1}{N}\left(1-\frac{k+1}{\sum_j(k_j+1)}\right)\nonumber\\
 &=&qm\frac{1}{N}-qm\frac{1}{N}\frac{k+1}{N+\sum_ik_i(i)}\nonumber\\
 &=&qm\frac{1}{t}-qm\frac{1}{t^2}\frac{k+1}{1+(1-q)2m+o(1)}\nonumber\\
\end{eqnarray}

\indent One thus has $\overline{A}=0$ and $\overline{B}=qm$. Since
$P(k)$ is power-law, by Lemma \ref{le:7} and
\begin{eqnarray}
 P(k)\sim Ck^{-(3-2q+\frac{1}{m})}
\end{eqnarray}
the scaling exponent is $3-2q+\frac{1}{m}$. So, the network is
scale-free.
\end{example}

\end{document}